\documentstyle[psfig,prd,aps]{revtex}
\def\L {\Lambda }
\def\l {\lambda } 
\def \t {\theta }

\def\a {\alpha }

\def \d {\delta }
\def \D {\Delta }

\def \g {\gamma }
\def \G {\Gamma }

\def \b {\beta }

\def \s {\sigma }
\def \e {\epsilon }
\def \ud { {1 \over 2} }

\def \capa {{\cal A }} 
\def \capb {{\cal B }} 
\def \capc {{\cal C }} 
\def \capd {{\cal D }}
\def \cala {{\cal A }}
\def \call {{\cal L }}
\def \cddd {{\cal D }}
\def \calf {{\cal F }}
\def \calr {{\cal R }}

\def \FLU {{ {\l ' }_{iJ'k} \l  _{iJk}^{'\star } \over (4\pi )^2 } }
\def \qslash {Q \kern-.5em\slash }
\def \pslash {p \kern-.5em\slash }
\def \ppslash {p' \kern-.5em\slash }
\def \Pslash {P \kern-.5em\slash }
\def \kslash {k \kern-.5em\slash }

\def \bea {\begin{equation}}
\def \eea {\end{equation}}

\def \ya  { X_1 }
\def \yb  { X_2 }
\def \yc  { X_3 }
\def \yd  { X_4 }
\def \ye  { X_5 }
\def \yf  { X_6 }
\def \yg  { X_7 }
\def \yh  { X_8 }

\def \pr  { Phys. Rev. }
\def \np { Nucl. Phys. }

\def \prl { Phys. Rev. Lett. } 
\def \pl { Phys. Lett. }

\begin{document}
\title{Polarized single top production at leptonic colliders from
broken R parity interactions incorporating CP violation \thanks {\it
Supported by the Laboratoire de la Direction des Sciences de la
Mati\`ere du Commissariat \`a l'Energie Atomique } }
\author{M. Chemtob and G. Moreau} \address{Service de Physique
Th\'eorique, CE-Saclay F-91191 Gif-sur-Yvette Cedex FRANCE}
\date{\today}

\maketitle
\begin{abstract}

The contribution from the R parity violating interaction, $\l '_{ijk}
L_i Q_j D_k^c $, in the associated production of a top quark
(antiquark) with a charm antiquark (quark) is examined for high energy
leptonic colliders.  We concentrate on the reaction, $l^-+l^+\to ( t
\bar c) + (c \bar t ) \to (b \bar l \nu \bar c) +(\bar b l \bar \nu c
) $, associated with the semileptonic top decay. A set of
characteristic dynamical distributions for the signal events is
evaluated and the results contrasted against those from the standard
model W-boson pair production background. The sensitivity to
parameters (R parity violating coupling constants and down-squark
mass) is studied at the energies of the CERN LEP-II collider and the
future linear colliders.  Next, we turn to a study of a CP-odd
observable, associated with the top spin, which leads to an asymmetry
in the energy distribution of the emitted charged leptons for the pair
of CP-conjugate final states, $b \bar l \nu \bar c$ and $\bar b l \bar
\nu c $.  A non vanishing asymmetry arises from a CP-odd phase,
embedded in the R parity violating coupling constants, through
interference terms between the R parity violating amplitudes at both
the tree and loop levels.  The one-loop amplitude is restricted to the
contributions from vertex corrections to the photon and Z-boson
exchange diagram. We predict unpolarized and polarized rate
asymmetries of order $ O(10 ^{-3}) - O(10 ^{-2}) $. An order of
magnitude enhancement may be possible, should the R parity violating
coupling constants $\l '_{ijk}$ exhibit a hierarchical structure in
the quarks and leptons generation spaces.
\end{abstract}
\vskip 1 cm 

{Saclay preprint t99/124; hep-ph/9910543; Phys. Rev. reference:
DK7265}

\section{Introduction}
\label{secintro}

The flavor non diagonal fermion-antifermion pair production,
$l^-l^+\to f_J \bar f_{J'},$ where $ J\ne J' $ are flavor labels,
represents a class of reactions where the high energy colliders could
contribute their own share in probing new physics incorporating flavor
changing and/or CP violation effects.  As is known, the standard model
contributions here are known to be exceedingly small, whereas
promising contributions are generally expected in the standard model
extensions.  (Consult ref.  \cite{chemtob} for a survey of the
literature.)  Of special interest is the case where a top quark
(antiquark) is produced in association with a lighter (charm or up)
antiquark (quark). The large top mass entails a top lifetime, $
\tau_{top} = [1.56 \ GeV ({m_t\over 180 \ GeV})^3]^{-1} $,
significantly shorter than the QCD hadronization time, $ 1/\L _{QCD}$,
which simplifies the task of jet reconstruction. \cite{toprev} The top
polarization effects also constitute a major attraction.
\cite{donoghue,topol0,kane,schmidt,chang1} The large top mass entails
a spin depolarization time of the top which is longer than its
lifetime, $\tau _{depol} = [1.7 MeV ({180\over m_t}) ]^{-1} >
\tau_{top} $, thus providing an easy access to top polarization
observables.  Polarization studies for the top-antitop pair production
reaction, in both production and decay, have been actively pursued in
recent years.  \cite{toppol,bartl,rigolin} (An extensive literature
can be consulted from these references.)

It appears worthwhile applying similar ideas to the flavor non
diagonal fermion pair production process involving a single top
production.  This reaction has motivated several theoretical studies
aimed at both leptonic ($ l^- l^+ , \ e \g $ and $\g \g $) and
hadronic ($ p \bar p , \ pp$) colliders.  Exploratory theoretical
studies have been pursued at an implicit level, via the consideration
of higher dimension contact interactions \cite{hikasa,han,abraham},
and at an explicit level, via the consideration of mechanisms
involving leptoquarks, \cite{barger} an extended Higgs doublet sector,
\cite{atwood,atwood1} supersymmetry based on the minimal
supersymmetric standard model with an approximately broken R parity,
\cite{chemtob,datta,oakes1,mahanta,litte} quark flavor mixing,
\cite{koide} standard model loops and four matter generations,
\cite{chang,huang99} or higher order standard model processes with
multiparticle final states, $ l^-l^+ \to t\bar c \nu \bar \nu
$. \cite{hou2} A survey of the current studies is provided in
ref.\cite{han}.

In this work, pursuing an effort started in our previous paper,
\cite{chemtob} we consider a test of the R parity violating (RPV)
interactions aimed at the top-charm associated production.  Our study
will focus on the contributions to the process, $ l^-l^+\to (t\bar c)
+(\bar t c) $, arising at the tree level from the trilinear RPV
interactions, $\l '_{ijk} L_i Q_j D_k^c $, via a $ \tilde d_{kR} $
squark exchange.  We examine the signal associated with the (electron
and muon) charged semileptonic decay channel of the top, $ t\to b W ^+
\to b l^+ \nu $.  The final states, $ (b l^+ \nu \bar c) + (\bar b l^-
\bar \nu c ), \ [l=e,\ \mu ] $ consist of an isolated energetic
charged lepton, accompanied by a pair of $ b $ and $ c $ quark
hadronic jets and missing energy.  The standard model background may
arise from the W-boson pair production reaction, $l^+ l ^-\to W^+W^-$,
and possibly, in the case of an imperfect $ b $ quark tagging, from
the $ b - \bar b $ quark pair production reaction, $l^+ l ^-\to b \bar
b $, followed by a semileptonic decay of one of the b quarks, $b \to c
l^- \bar \nu $.

The present work consists of two main parts. In the first part, we
discuss the signal associated with the top semileptonic decay channel.
We evaluate a set of characteristic dynamical distributions for the
signal and for the standard model background and obtain predictions
for the effective rates based on a judicious choice of selection cuts
on the final state kinematical variables.  Our discussion will develop
along similar lines as in a recent work of Han and Hewett, \cite{han}
which was focused on the contributions initiated by the dimension, $
\cddd =6$, four fields couplings of the Z-boson with fermion pairs and
the neutral Higgs boson.  In the second part of the paper, we examine
a specific CP-odd top polarization observable which corresponds to an
asymmetry in the energy distribution of the final state charged lepton
with respect to the sign of its electric charge.

The contents of the paper are organised into 3 sections.  In Section
\ref{sec2}, we focus on the total and partial semileptonic decay rates
for both the signal and standard model background, allowing for the
case of an imperfect $ b $ quark tagging.  We discuss the constraints
from the indirect bounds on the RPV coupling constants, study the
dependence of rates on the down-squark mass parameter and evaluate a
set relevant dynamical distributions that are of use in devising an
appropriate set of selection cuts.  In Section \ref{sec3}, we discuss
a test of CP violation involving top polarization effects. The CP
violating observable arises through interference terms between the
tree and one-loop contributions to the amplitude and a CP-odd phase
which is embodied in the RPV coupling constants.  Following an
approach similar to one used in earlier proposals,
\cite{schmidt,chang1} we describe the top production and decay by
means of a factorization approximation and examine the induced charge
asymmetry in the energy distribution of the final state charged
leptons. The production amplitudes are evaluated in the helicity
basis.  Our main conclusions are summarized in Section \ref{sec4}.

\section{Top-charm associated production}
\label{sec2}
\subsection{ Integrated rates}

\begin{figure} [h]
\begin{center}
\leavevmode \psfig{figure=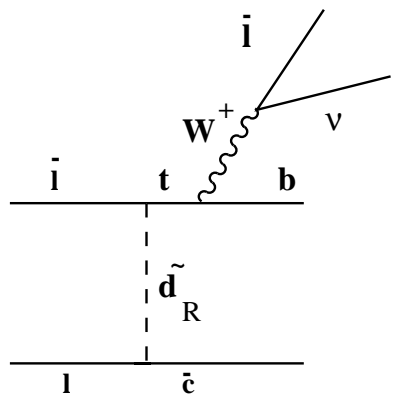}
\end{center}
\caption{Feynman diagram for the tree level amplitude of the process,
$ l^+l^- \to \bar c t \to \bar c b l^+ \nu $.}
\label{fig1}
\end{figure}

In a $ l^- l^+$ collision, the tree level transition amplitude for
single top production, as initiated by the RPV interactions, $ \l
'_{ijk} L_iQ_jD_k^c $, proceeds via the $u$-channel exchange of a
right-handed down-squark, $\tilde d_{kR}$, as represented in
Fig.\ref{fig1}.  By use of a Fierz ordering identity, the transition
amplitude for the flavor non diagonal production of an up
quark-antiquark pair, $ l^- (k) + l^+ (k') \to u_J (p) +\bar u_{J'}
(p') $, can be written in the form of a Lorentz covariant vectorial
coupling,
\begin{eqnarray}
M^{JJ'}_{t}= -{\l ^{'\star }_{lJk} \l '_{lJ'k} \over 2(u-m^2_{\tilde
d_{kR} }) } \bar v_L(k') \g^\mu u_L(k)\bar u_L(p)\g_\mu v_L(p').
\label{eq1pp}
\end{eqnarray}
We shall specialize henceforth to the case of electron-positron
colliders, corresponding to the choice $ l=1$ for the generation
index.  The squared amplitude, summed over the initial and final
fermion spins, reads: \cite{chemtob}
\begin{eqnarray}
\sum_{pol} \vert M^{JJ'}_t \vert ^2 &=& N_c \bigg \vert -{ \l'_{1J'k}
\l {'^\star } _{1Jk} \over 2(u-m^2_{\tilde d_{kR} } )} \bigg \vert ^2
16 (k\cdot p')(k'\cdot p ) .
\label{equ1p}
\end{eqnarray}
The production rate for unpolarized initial leptons, integrated over
the scattering angle in the interval, $ 0\le \vert \cos \t \vert \le
x_c$, is given by the analytic formula,
\begin{eqnarray}
\s &= &{ N_c \vert \l'_{1J'k} \l^{'\star } _{1Jk} \vert ^2 \over 64
\pi s^{2} } [(u_--u_+) +(2\tilde m^2-m_J^2 -m_{J'} ^2 )\ln \vert
{u_--\tilde m^2 \over u_+-\tilde m^2} \vert \cr & -& (\tilde
m^2-m_J^2)(\tilde m^2-m_{J'}^2) ({1\over u_--\tilde m^2 } -{1\over
u_+-\tilde m^2 } ) ],
\label{equ1q}
\end{eqnarray}
where, $ u_\pm = m_J^2 -\sqrt s (E_p \pm p x_c )$.  For the top-charm
associated production case, in the limit, $ m_J = m_t >> m_{J'} = m_c
$, one has, $ u_+ \simeq m_t^2 -s, \ u_- \simeq 0$.  For fully
polarized initial beams, since the RPV amplitude selects a single
helicity configuration for the initial state leptons, $l^-_L l^+_R$,
(left handed $l^-$ and right handed $ l^+$) the corresponding
polarized rate would be still described by the same formula as above,
only with an extra enhancement factor of $4$.  The predicted rates for
$ t\bar c $ production are controlled by quadratic products of the RPV
coupling constants, $\l ^{'\star }_{13k} \l '_{12k},\ [ k=1,2,3]$ and
the squark mass, $m_{\tilde d_{kR} }$, denoted for short as, $\tilde
m$. Allowing for the existence in the RPV interactions of an up-quark
flavor mixing, such as would be induced by the transformation from
flavor to mass basis, one may express the amplitude in terms of a
single RPV coupling constant and the CKM matrix, $V$, by rewriting the
coupling constant dependence as, $ \l'_{12k} \l {'^\star } _{13k} \to
\l _{1Mk}^{'\star } \l '_{1M'k} (V^\dagger )_{M'2}(V^\dagger ) ^\star
_{M3} $, and selecting the maximal contribution associated with the
configurations, $M=M'=2 $ or $ 3$.  This yields the order of magnitude
estimate, $\l _{13 k} ^{'\star } \l '_{12 k} \to \vert \l '_{12k}\vert
^2 (V ^\dagger )_{22} (V^\dagger )^\star _{23} \approx 2 \vert \l
'_{12k}\vert ^2 \l^2 $ or $ 2 \vert \l '_{13k}\vert ^2 \l^2 $,
respectively, where, $\l \approx \sin \t_c \approx 0.22 $, denotes the
Cabibbo angle parameter.

We pause briefly to recall the current bounds on the RPV coupling
constants of interest in the present study. \cite{reviews} The
relevant single coupling constant bounds are, $ \l '_{12k} < 4.
\times 10^{-2} ,\ \l '_{13k} < 0.37 $ (charged current universality);
$ \l '_{1j1} < 3 \times 10^{-2} $ (atomic physics parity violation); $
\l '_{12k} < 0.3-0.4, \ \l '_{13k} < 0.3 \ - \ 0.6 $ (neutral current
universality); and $\l '_{122} < 7. \times 10^{-2},\ \l '_{133} < 3.5
\times 10^{-3} $ (neutrino Majorana mass).  \cite{godbole} The
superpartners scalar particles masses are set at $ 100 \ GeV$. Unless
otherwise stated, all the dummy flavor indices for quarks and leptons
are understood to run over the three generations.  Using the above
results for individual coupling constants bounds, we may deduce for
the following upper bounds on the relevant quadratic products,
\cite{grossman} $ \l '_{13k} \l '_{12k} < [O(10^{-3}), \ O(10^{-2}), \
O(10^{-4})] , [k=1,2,3]$.  The indirect quadratic products bounds, $
\l '_{ijk} \l '_{i' 3k} < 1.1 \times 10^{-3} , \ \l '_{ijk} \l '_{i' j
3} < 1.1 \times 10^{-3} , \ [ i' =1,2 ] $ ($B \to X_q \nu \bar \nu $)
are roughly comparable to these single coupling constants bounds.  We
also note that using the CKM flavor mixing along with a single
dominant coupling constant in the current basis, as described at the
end of the previous paragraph, may not be especially beneficial in
avoiding the above stronger pair product bound.  The bound on the
corresponding coupling constant factor, $ 2 \vert \l '_{13k}\vert ^2
\l^2 < O(10^{-2}) $, is competitive for the generation indices, $
k=1,3$.

Numerical results for the integrated rates have already been reported
in previous works \cite{mahanta,chemtob}.  Setting the relevant RPV
coupling constants at the reference value, $\l '= 0.1 $, one predicts
rates of order $ 1 \ - \ 10 \ fb$, for $ \tilde m = O (100) \ GeV$.
As the center of mass energy varies in the interval, $\sqrt s = 192 \
- \ 1000 \ GeV$, the rates rise sharply from threshold, reaching
smoothly a plateau around $ \sqrt s \simeq 400 \ GeV$. This contrasts
with the predictions from gauge boson mediated higher dimension
interactions \cite{han} where the rise of the rates with incident
energy is a more gradual one.  The rates are also found to have a
strong dependence on $\tilde m$, which weakens for increasing center
of mass energies.  One may roughly parametrize the dependence on $ s $
and $ \tilde m$ by the approximate scaling law, $\s \approx ({\l ' \l
' \over 0.01 } )^2 ({100 \ GeV \over \tilde m })^{x (s)},$ where the
power exponent is a fastly decreasing function of energy, taking the
approximate values, $ x (s) \approx [3.65, \ 1.86, \ 0.94] $, at,
$\sqrt s =[0. 192, \ 0.5 , \ 1. ] \ TeV$.

\begin{figure} [h]
\begin{center}
\leavevmode \psfig{figure=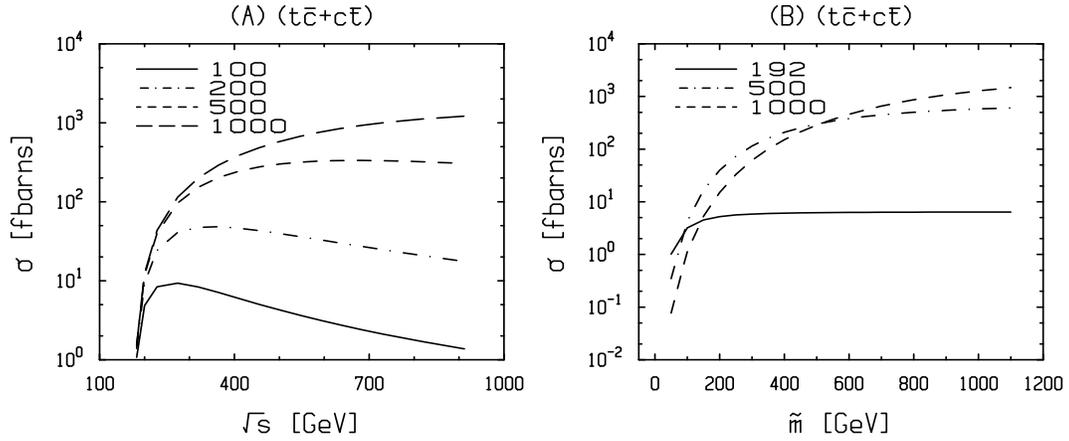,width=14cm}
\end{center}
\caption{The total integrated rate for the RPV induced reaction, $
l^+l^- \to( t \bar c) +(\bar t c) $, setting the values of the
relevant coupling constants as, $ \l '_{12k} = \l '_{ 13k} =0.1 \ (
\tilde m /100 \ GeV ) $, is plotted in window $(A)$ as a function of
center of mass energy, $ s^{1/2} $ for fixed down squark mass, $
\tilde m =[ 100, \ 200, \ 500, \ 1000 ] \ GeV $ and in window $(B)$ as
a function of $\tilde m$ for fixed $s^\ud =[ 192, \ 500, \ 1000 ] \
GeV $. We integrate over an interval of the scattering angle, $ 0\le
\vert \cos \t \vert \le 0.9848 $, corresponding to an opening angle
with respect to the beams axis larger than $10^{o } $. }
\label{fig2}
\end{figure}

Although the predicted rates seem to be severely constrained by the
above indirect bounds, one could envisage an optimistic scenario where
the supersymmetry decoupling limit, $\tilde m \to \infty $, is
realized with fixed values for the products, $ \l '_{ijk} ({100 \
GeV\over \tilde m }) \approx 0.1 $, consistently with the current
indirect bounds.  The results obtained with this prescription are
displayed in Figure \ref{fig2}.  The integrated rates now depend on $
\tilde m $ as, $ \s \propto ({100 \ GeV \over \tilde m })^{-4+x (s)}$,
which leads at high energies to an enhancement by up to three orders
of magnitudes, compared to the case where the RPV coupling constants
are taken independent of $\tilde m$. The initial energy of LEP-II
falls right in the regime where the cross section is sharply rising
with increasing initial energy.  The decrease with increasing $\tilde
m$ is stronger at LEP-II energies than at the future linear colliders
energies.  Note that at the largest values of the superpartner mass,
$\tilde m \simeq 1.  \ TeV$, the RPV coupling constants in our
prescription enter a strong coupling regime ($\l ' =O( 1) $) and it is
not clear then whether the tree level prediction makes sense.

Next, we consider the process incorporating the top semileptonic
decay, as pictured by the Feynman diagram shown in Fig.\ref{fig1}. We
assume that the top decay is dominated by the electroweak semileptonic
decay channel, with branching fraction, $ B (t\to b +W^+) \approx
1$. We also include the pair of CP-conjugate final states, $t\bar c $
and $c \bar t$ production, which multiplies the rate by a factor of
$2$. Note, however, that we restrict ourselves to the $ u_{J'} = c$
charm quark mode only.  The numerical results for rates, including a
branching fraction factor of ${2\over 9} $ (experimental value, $
21.1\% $) to account for the $W \to l \nu , \ [l= e, \ \mu ]$ decay
channels, are displayed in Table \ref{table1}.  We also show the
standard model background rate from the $W $-boson pair production,
$l^-l^+\to (W^+ W^-) \to ( \ l^+ \nu \bar u_i d_j ) + ( l^- \bar \nu
\bar d_j u_i) $, with one $ W$-boson decay leptonically and the other
hadronically, where $ i, \ j $ are generation indices.  The
irreducible background from, $ W ^- \to \bar c b $ or $ W ^+ \to \bar
b c$, is strongly suppressed, due to the small branching factor, given
approximately by, $ 0.32 \vert V_{cb} \vert ^2 \approx 5 \ 10 ^{-4}$.
It is safer, however, to allow for the possibility where the light
quark hadronic jets could be misidentified as $ b $ quark hadronic
jets.  Accounting for the leptonic decay for one of the $W$-boson and
the hadronic decay for the other $ W$-boson, introduces for the total
rate, which includes all the subprocesses, the branching fraction
factor, $ 2\times (21.1 \pm 0.64 )\% \times (67.8\pm 1.0)\% = 0.286
\pm 0.024$.  Our numerical results in Table \ref{table1} for the
standard model background rates are in qualitative agreement with
those quoted ($ \s = [2252, \ 864] \ fb$ at $ s^\ud = [0.5, \ 1. ] \
TeV$) by Han and Hewett.  \cite{han} One should be aware of the
existence of large loop corrections to the $ W^+ W^-$ production rate,
especially at high energies. The predictions including the electroweak
and QCD standard model one-loop contributions read, \cite{beenakker}
$\s = [4624, \ 1647, \ 596 ] \ fb$ at $ s^\ud = [0.192, \ 0.5, \ 1. ]
\ TeV$.  We conclude therefore that our use of the tree level
predictions for the $ W^+W^-$ background overestimates the true cross
sections by $ [9\%, 20 \%, 32 \%] $ at the three indicated energies.

\begin{table} 
\begin{center}
\vskip 1 cm
\caption{ Production rates for the top-charm production signal and the
W-boson pair production background.  The line entries give
successively the total integrated rate for the reaction, $l^+l^- \to (
t \bar c) + (c \bar t )$, using, $\l '= 0.1, \ \tilde m =100 \ GeV$,
the rate for signal events, $(b \bar l \nu \bar c) +(\bar b l \bar \nu
c ) $, associated with the top semileptonic decay, the W-boson pair
production background rate, $l^-l^+\to W^+ W^-\to (l^+ \nu \bar u_i
d_j ) + (l^-\bar \nu \bar d_j u_i ) $, and the corresponding cut
signal and background rates, as obtained by applying the selection
cuts quoted in the text.  The results include the first two
generations of charged leptons, $ \ l=e, \ \mu $. }
\vskip 0.5cm
\begin{tabular}{|c|ccc|}
\tableline $ Energy (TeV) $ & $ 0.192 $& $0.5 $ & $1.0 $ \\ \tableline
Total rate $ \s (fb) $ & $ 4.099 $& $ 4.291 $ & $1.148$ \\ \tableline
Signal (fb) & $ 0.68 $ & $ 0.91 $ & $ 0.24 $ \\ \tableline $W^+W^-$
Background (fb) & $ 5076$ & $ 2080 $ & $ 876 $ \\ \tableline Cut
Signal (fb) & $ 0.54 $ & $ 0.74 $ & $ 0.21$ \\ \tableline $W^+W^-$ Cut
Background (fb) & $ 17. $ & $5.0 $ & $ 2.6 $ \\ \tableline
\end{tabular}
\end{center}
\label{table1}
\end{table}

Let us discuss briefly other possible sources of background. The next
important contribution is that arising from the non resonant W-boson
virtual propagation in the amplitude with the intermediate $ W^+ W^-$
bosons branching into four fermions ($l \nu q\bar q'$).  This could be
possibly estimated by subtracting the resonant contribution from the
total background cross sections, weighted by the suitable branching
factors, as independently evaluated by numerical methods in the
literature.  The results for the integrated total cross sections, $
l^-l^+ \to (4 f) + (4 f + \g )$, \cite{bardin} including the initial
state radiation and Coulomb corrections, indicate that the off-shell
contributions amount to a small relative correction lower than $
O(10\% ) $.  Alternatively, one may consider, after reconstructing the
neutrino momentum from the missing energy, a procedure to impose
suitable cuts on the $ bW$ invariant mass, aimed at suppressing the
non resonant production background.

 One other potentially important background is that arising from the
$b- \bar b$ quark pair production reaction, $l^+l^- \to \g ^\star / Z
\to b \bar b \to \bar b ( cl^-\bar \nu ) + b (\bar cl^+\nu ) $.
\cite{mahanta} The numerically derived predictions for the rates, as
obtained by means of the PYTHIA generator, are: $ \s = [1.631 \ 10^4 ,
\ 2.12 \ 10^ 3 , \ 5.35 \ 10^2 ] \ fb$, at $ s^\ud = [0.192, \ 0.5, \
1. ] TeV$.  It would appear desirable, in view of these large
predicted rates, to eliminate this background by performing a double $
b $ quark tagging analysis on the events sample. This can be performed
at a reasonably low cost, given that the detection efficiency of $b$
quark jets is currently set at $ 50\%$.  If one performs a single $ b
$ quark tagging, the rates for the corresponding events, $l^+l^- \to
\g ^\star / Z \to b \bar b \to ( cl^-\bar \nu ) \bar b $, are reduced
by a branching fraction, $ B(b \to c l \nu ) \approx 10\% $, but this
is compensated by the probability of misidentifying a light quark jet
as a $ b $ quark jet, which lies at the small value of $ 0.4 \% $ with
the current silicon vertex techniques.  If no $ b $ quark tagging is
performed at all, then the above large rates may make it necessary to
resort to an analysis of isolation cuts of the type to be discussed in
the next subsection.

\subsection{Distributions  for the semileptonic top decay events}
\label{sec21} 
\begin{figure} [t]
\centerline{ \psfig{figure=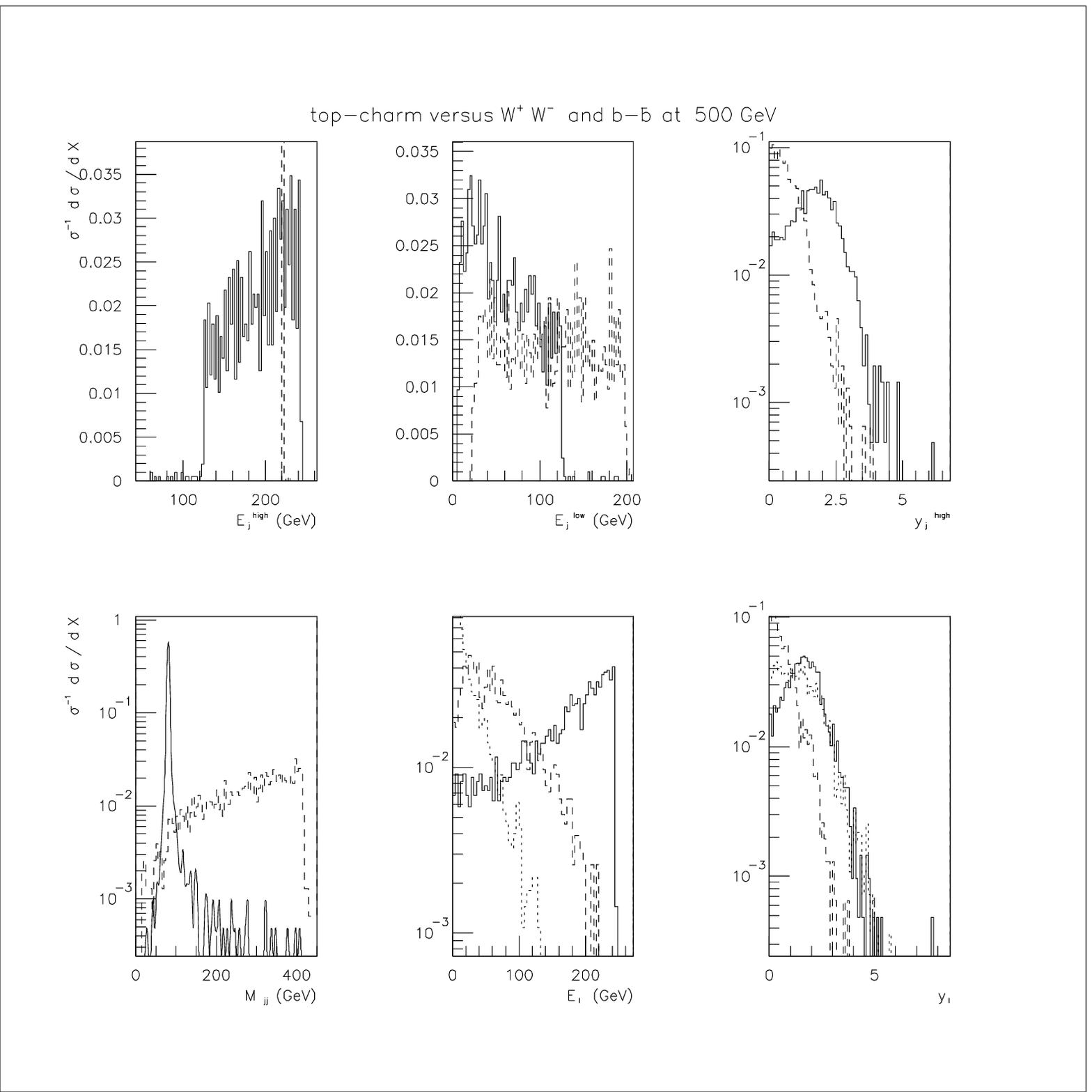} }
\vskip 1 cm
\caption{ Normalized dynamical distributions associated with the
signal events, $l^+l^-\to (t \bar c) +(\bar t c) \to (b l^+ \nu \bar
c) +( \bar b l^-\bar \nu c)$, (dashed line) at $\tilde m =100 \ GeV$,
and the background events, $ l^+l^-\to W^+W^- \to ( l^+ \nu \bar q q'
) + ( l^- \bar \nu q\bar q' ) $, (continuous line) at a center of mass
energy, $ s^\ud = 500 \ GeV$. The kinematical variables in the
histograms, from left to right and up to down, are the jets maximum
energy, the jets minimum energy, the rapidity for the highest energy
jet, the di-jet invariant mass, the charged lepton energy and the
charged lepton rapidity.  The charged lepton energy and rapidity
distributions are also plotted for the $ b-\bar b$ background
production events, $ l^+l^-\to b \bar b \to l^\pm + hadrons $ (dotted
line).  }
\label{figtopp}
\end{figure}

In order to separate the signal from background, we consider the same
set of characteristic final state kinematical variables as proposed in
the study by Han and Hewett. \cite{han} These are the maximum and
minimum energy of the two jets, $ E_j^{high}, \ E_j^ {low}$, the dijet
invariant mass, $ M_{jj} $, the charged lepton energy, $E_l,$ and
rapidity, $y_l = \ud \log {E_l +p_{l\parallel } \over E_l-p_{l
\parallel } } $.  The distributions in these $6$ variables for the
signal and background, at a center of mass energy, $ \sqrt s = 0.5
TeV$, are plotted in Fig.\ref{figtopp}.  These numerical results were
obtained by means of the PYTHIA \cite{sjostrand} event generator. One
notices marked differences between signal and background.  The maximum
jet energy distribution is uniformly distributed for the background
but sharply peaked for the signal, where the peak position is
determined by the top mass and the incident energy as, $m_t^2 = (m_c^2
-s +2\sqrt s E_p)$.  The minimum jet energy is uniformly distributed
for both signal and background, but happily the corresponding
intervals are very partially overlapping.  The signal events rapidity
distributions for the maximum energy jet are more central for signal
than background. A similar trend holds for the lepton rapidity
distributions.  The dijet invariant mass is a most significant
variable in discriminating against the background due to its
pronounced peak at the $ W $ mass. For the signal, the dijet invariant
mass is uniformly spread out. Although we do not show here the
distributions for the top mass reconstruction, this also features a
strong contrast between a strongly peaked signal and a uniform
background. The lepton energy distributions for the signal and
background are peaked at the opposite low and high energy ends of the
physical interval, respectively. This is a familiar effect associated
with the correlation between the W-boson spin polarization, which is
predominantly longitudinal in the top decay and transverse in the
direct W-boson decay, and the velocity of the emitted charged lepton.
In the signal decay amplitude, $ t \to b\bar l \nu $, the fact that
the left handed b-quark must carry the top polarization, forces the
lepton to travel with opposite velocity to that of top.  In the
background decay amplitude, $ W^-\to l \bar \nu $, the charged lepton
is emitted with a velocity pointing in the same direction as that of
the W-boson. Thus, the Lorentz boost effects on the emitted charged
leptons act in opposite ways for the signal and background events.

\begin{figure} [t]
\centerline{ \psfig{figure=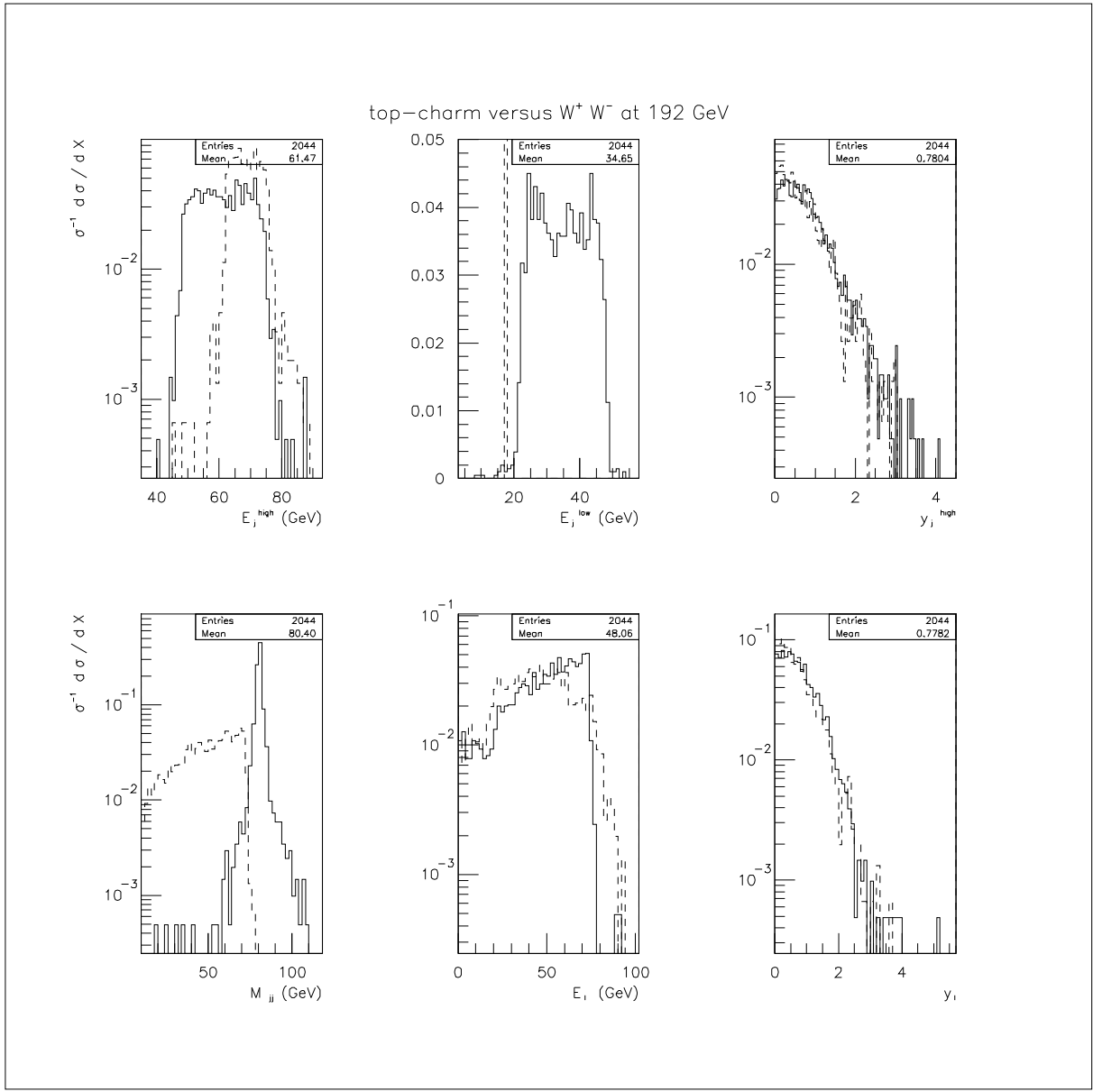} }
\vskip 1 cm
\caption{ Same distributions as in Fig.\ref{figtopp}, at a center of
mass energy $ s^\ud = 192 \ GeV$. }
\label{figtopp192}
\end{figure}

While the above distinctive features between signal and background
events get further pronounced with increasing center of mass energy,
opposite trends occur as the initial energy is lowered.  The
distributions at the LEP-II center of mass energy, $ \sqrt s = 0.192 \
TeV,$ are plotted in Fig.\ref{figtopp192}.  At this energy, the
monovalued distribution for the signal jet, which is now the softer
lower energy jet, is still well separated from the corresponding
background jet distribution.  So, this variable, along with the dijet
invariant mass stand up as useful discrimation tests for the signal.
By contrast, the energy and rapidity distributions for the maximum
signal jet may not be easily distinguished from the
background. Similarly, the lepton energy distributions in the signal
and background are overlapping due to the small Lorentz boost effect.

The distributions obtained with the RPV interactions are rather
similar to those found with the higher dimension operator
mechanism. \cite{han} This is due to the formal structure of the RPV
amplitude, involving an effective $u$-channel vector particle
exchange.  In fact, the selection cuts proposed by Han and Hewett
\cite{han} appear to be quite appropriate also in the RPV case, and,
for convenience, we recapitulate below the cut conditions used to
characterize the selected events.

$ E_j^{low} < 20,\ E_j^ {high} > 60,\ E_l > 0,\ \d_{jj}> 10,\ \d_t <
5, \quad [ \sqrt s = 192] $

$ E_j^{low} > 20,\ E_j^ {high} > 200,\ E_l < 150, \ \d_{jj}> 10,\ \d_t
< 40, \quad [\sqrt s = 500 ] $

$ E_j^{low} > 20,\ E_j^ {high} > 460,\ E_l< 350,\ \d_{jj} > 10,\ \d_t
< 100.  \quad [\sqrt s = 1000] $

 The above listed variables correspond to the minimum and maximum
energy of the two jets, $ E_j^ {low} , \ E_j^{high}, $ the charged
lepton energy, $E_l, $ the distance between the dijet invariant mass
and W-boson mass, $\d_{jj} = \vert M_{jj} -m_W \vert, $ the distance
of the reconstructed top mass to the true mass, $ \d_t = \vert
m_t^{reconst}-m_t \vert $.  The assigned numerical values are all
expressed in GeV units.  Besides the above cuts, we also impose the
usual detection cuts on energies and rapidities, $ E_{j,l} > 10 \ GeV,
\ \vert \eta_{j,l} \vert < 2$, aimed at removing the particles
travelling too close to the beam pipe.  We allow for the detection
efficiency of the particle energies only in an approximate way,
namely, by accounting for the following approximate uncertainties, $\D
E /E = 40 \%, \ 10 \%$, on the jets and lepton energies, respectively,
at the level of imposing the above selection cuts, rather than by the
usual procedure of performing a Gaussian smearing of the particle
energies.
 
The numerically evaluated efficiencies on the signal and background
events are, $ \e_S \simeq 0.8 , \ \e_B \simeq 3. 10^{-3} $, with a
very weak dependence on the center of mass energy and, for the signal,
a weak dependence on the mass parameter $\tilde m$, which was set at $
\tilde m = 100 \ GeV$ in the numerical simulations.  After applying
the cuts, the background rates are, $ \s_B \e_B = [ 17., \ 5., \ 2. ]
\ fb, $ and the signal rates, $ \s_S\e_S = [ 0.68, \ 0.74, \ 0.21 ] \
fb $, for $\sqrt s = [192., \ 500., \ 1000 ] \ GeV$.  The results for
the cut signal and background rates, as given in Table \ref{table1},
show that the background is very significantly reduced by the cuts.
The situation is clearly far more favorable for the future linear
colliders than for LEP-II.  Nevertheless, the number of surviving
signal events is still one order of magnitude below that of the
surviving background, so that the option of cutting down the
background by means of a $b $ quark tagging procedure is to be
preferred since the ensuing reduction would be much more drastic. An
integrated luminosity of $\call = 100 \ fb^{-1}$ would lead to a
number of signal events, $(\l '_{12k} \l'_{13k} /10^{-2} ) \times
O(30) $.

We have also performed an indicative event generator study of the
background, $ l^+l^-\to b \bar b \to l^\pm + hadrons $, restricting
consideration to the emitted charged leptons only. A jet
reconstruction of the partonic level distributions is a task beyond
the scope of the present work. We focus on the first charged lepton
emitted during the semileptonic decays of the produced $B , \ \bar B $
mesons, since this carries the largest velocity. As seen on
Fig.\ref{figtopp}, the distribution for the first emitted charged
lepton energy is peaked at low energies.  One expects that the most
energetic lepton is that produced in the semileptonic decays of the $B
$ mesons.  The rapidity distribution is less central than for the
signal and nearly overlaps with that of the $ W^+W^-$ background.
Therefore, imposing the additional lower bound cut on the lepton
energy, say, at $ E_l > 20 \ GeV$, for $s^\ud = 500 GeV$, should be
sufficient to appreciably suppress the $ b-\bar b$ background without
much affecting the signal.

\begin{figure} [h]
\centerline{ \psfig{figure=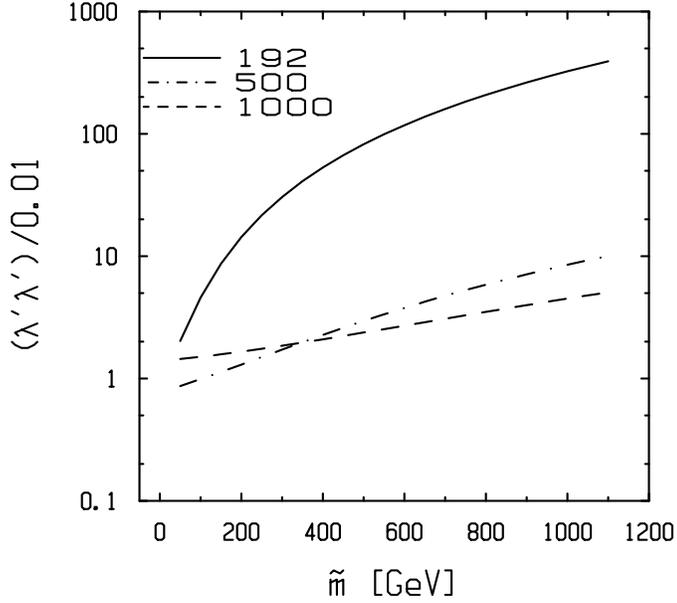}}
\vskip 1.  cm
\caption{ Sensitivity reach plot for the RPV coupling constants
product, $\l '_{12k} \l '_{13k} /0.01$, as a function of the down
squark mass, $\tilde m$, for fixed center of mass energy, $ s^\ud =[
192. \ , 500 , \ 1000 ] \ GeV$, and corresponding fixed integrated
luminosity, $ \call = [2., \ 100., \ 100. ]\ fb^{-1} $, using an
acceptancy for the background, $ \e _B = 3\ 10 ^{-3} $, and an
acceptancy for the signal, $ \ \e_S = 0.8 $, assumed to be independent
of $\tilde m$.}
\label{figsens}
\end{figure}

We may infer the reach with respect to the free parameters by
evaluating the statistical significance ratio for a discovery, as
defined by, $ \hat \sigma = {S\over \sqrt { S+B} } ,\ S = \s_{S} \call
,\ B =\s_{B} \call $, where $ \call $ denotes the integrated
luminosity.  Setting this at the value, $ \hat \sigma = 3 $,
corresponding to a $ 95 \% $ confidence level, one deduces a
dependence of the RPV coupling constant as a function of the
superpartner mass parameters for a fixed initial energy and integrated
luminosity.  The sensitivity reach contour plot for the relevant
parameters, $ \l ' \l '= \l '_{12k} \l _{13k} ^{'\star } $ and $\tilde
m = m_{ \tilde d _{kR} } $, is shown in Fig.\ref{figsens}.  We note
that the sensitivity limit on the product of coupling constants, $ \l
' \l '$, scales with the luminosity approximately as, $ 1/ \sqrt \call
$.  While the reach on the RPV coupling constants products, $\l
'_{12k} \l '_{13k} < O(10^{-1}) $, lies well above the current
indirect bounds, this covers a wide interval of the down squark mass
which extends out to $ 1 TeV$.  To compare with analogous collider
physics processes, we note that while the flavor diagonal fermion pair
production reactions, $ e^-e^+ \to f_J\bar f_J$, may have a higher
sensitivity reach, these are limited to information on the single
coupling constants, $ \l '_{1jk}$ \cite{choudhee}.  The special
reaction, $ e^-e^+ \to b\bar b $, proceeding via a sneutrino
$s$-channel resonance, may probe quadratic products such as, $
\l_{131} \l '_{333} $, \cite{erler} or $ \l_{131} \l '_{311}$
\cite{kalitev} at levels of $ O(10^{-3})$, but this is subject to the
existence of a wide sneutrino resonance.  The $t\bar b$ associated
production at the hadronic Fermilab Tevatron \cite{datta,oakes1} and
the Cern LHC \cite{oakes1} colliders can be initiated via a charged
slepton $ \tilde e_{iL} $ $s$-channel exchange. The sensitivity reach
on the linear combination of quadratic coupling constants products,
$\l'_{i11} \l'_{i33} $, is of order $ 10^{-2} - 10^{-1}$. This
information should prove complementary to that supplied by our study
aimed at the leptonic colliders.  To conclude this brief comparison,
we observe that the information provided by the single top production
reaction appears to be rather unique in view of the very
characteristic signature of the associated events.

\section{Top polarization observables and a test of  CP violation}
\label{sec3}

Should single top production become experimentally observable in the
future, an important next step to take is in examining top
polarization observables.  In this section, we present an approximate
study for the top semileptonic decay signal in top-charm associated
production aiming at a test of CP violation.  We exploit an idea which
was developed in early studies of $ t-\bar t$
production. \cite{kane,schmidt} Interesting extensions are currently
pursued \cite{toppol,bartl,rigolin}. The basic observation is that any
CP-odd quantity depending on the top polarization, such as the
difference of rates between the pair of CP conjugate reactions, $ \s (
l^- l^+ \to t_L \bar c ) - \s ( l^- l^+ \to \bar t_R c ) $, can become
observable by analyzing the top polarization through the kinematical
distributions of its emitted decay ($b$ quark or charged lepton)
products.  An especially interesting observable is the charged lepton
energy distribution for a polarized top.  Any finite contribution to
the CP-odd observables must arise through an interference term
involving imaginary parts of loop and tree amplitudes factors, the
loop amplitude factor bringing a CP-even final state interaction
complex phase with the CP-odd relative complex phase arising from the
coupling constants in the product of loop and tree amplitudes.

\subsection{Helicity basis amplitudes} 
\label{sec30}
 Building on our previous work, \cite{chemtob} we shall combine the
tree level RPV induced amplitude discussed in Section \ref{sec2} with
the one-loop RPV induced amplitude associated to the photon and
Z-boson exchange diagrams, restricting ourselves to the vertex
corrections in the electroweak neutral current vertices, $ \g \bar f
_J (p) f_{J'} (p') $ and $ \ Z \bar f _J (p) f_{J'} (p') $.  The
Z-boson vertex admits the general Lorentz covariant decomposition,
\begin{eqnarray}
J_\mu ^{ Z }& = & - {g \over 2 \cos \t_ W} \G ^{JJ'} _\mu (Z), \cr \G
^{JJ'} _\mu (Z) & =& \g_\mu ( A ^{JJ'}_L(f) P_L + A ^{JJ'}_ R (f) P_R)
+{1\over m_J +m_{J'}} \s_{\mu \nu } (p+p')^\nu (i a ^{JJ'} +\g_5 d
^{JJ'} ),
\label{equ321}
\end{eqnarray}
where the vectorial vertex functions, $ A^{JJ'} _{L , R} = A^{JJ'} _{L
, R}\vert_{tree} + A^{JJ'} _{L , R}\vert_{loop}$, have a tree level
contribution given by, $ A_{L,R} ^{ JJ' }\vert _{ tree} = \d_{JJ'}
a_{L,R} (f), \ a_{L,R} (f) = 2T_3^{L,R} -2Q(f) \sin ^2 \t_W $, and the
tensorial vertex functions, $ a^{JJ'},\ d^{JJ'} $, are associated with
the anomalous transition magnetic moment and the CP-odd, P-odd
electric transition dipole moment, respectively.  An analogous
decomposition applies for the photon, $J_\mu ^{\g } = -{g \sin \t_ W
\over 2 } \G ^{JJ'}_\mu ( \g ),$ with $ a_{L,R} (f) =2Q(f) $,
determined by the electric charge $Q(f)$.  It is convenient to work
with the $ Z f _J\bar f_{J'}$ vertex in the alternate Lorentz
covariant decomposition, $\G ^{JJ'} _\mu (Z) =\g_\mu (\capa -\capb
\g_5 ) + \ud (p-p')_\mu (\capc -\capd \g_5 ),$ where the vertex
functions, $\capa , \ \capb , \ \capc , \ \capd $, (omitting the up
quarks generation indices $ J, \ J'$ for convenience) are related to
the previously defined vectorial and tensorial ones,
eq. (\ref{equ321}), as,
\begin{eqnarray}
\capa & =& {1\over 2} (A ^{JJ'}_L(f) +A ^{JJ'}_R(f) ) +a ^{JJ'} , \
\capb = {1\over 2} (A ^{JJ'}_L (f) -A ^{JJ'}_R (f) ) + { m_J -m_{J'}
\over m_J +m_{J'} }i d ^{JJ'} , \cr \capc & =& -{2 \over m_J +m_{J'} }
a ^{JJ'} , \ \capd = -{2 \over m_J +m_{J'} } i d ^{JJ'}.
\label{equ51}
\end{eqnarray} 
The one-loop Z-boson exchange amplitude may then be written in the
form,
\begin{eqnarray}
M^{JJ'}_l (Z) & = &\bigg ({g\over 2 \cos \t_W}\bigg )^2 \bar v (\vec
k',\mu ')\g_\s \bigg (a(e_L) P_L+a(e_R)P_R \bigg )u(\vec k, \mu )
{1\over s-m_Z^2+im_Z\G_Z} \cr \times &&\bar u (\vec p,\l ) [\g^\s
(\capa -\capb \g_5 ) + \ud (p-p')^\s (\capc -\capd \g_5 ) ] v({\vec p
' }, \l ') .
\label{equ322}
\end{eqnarray}
Combining the above loop amplitude with the RPV tree amplitude,
eq. (\ref{eq1pp}), which we rewrite as,
\begin{eqnarray}
M ^{JJ'} _t = \calr \bar v \g_\mu (1-\g_5) u \bar u \g^\mu (1-\g_5) v,
\quad \calr= -{ \l _{1Jk}^{'\star } \l '_{1 J'k} \over 8
(u-m^2_{\tilde d_{kR } } ) } ,
\label{equ323}
\end{eqnarray}
one obtains,
\begin{eqnarray} 
M^{JJ'}&=& M^{JJ'} _t + M^{JJ'} _l(Z) = [ ( G a^+ \capa + \calr )( \g
_\mu )( \g ^\mu ) - ( G a^+ \capb + \calr )( \g _\mu )( \g ^\mu \g_5)
\cr & - & ( G a^- \capa + \calr )( \g _\mu \g_5)( \g ^\mu ) + ( G a^-
\capb + \calr )( \g _\mu \g_5 )( \g ^\mu \g_5) \cr & + & \ud
(p-p')^\mu [ G a^+ \capc ( \g _\mu )( 1) - G a^+ \capd ( \g _\mu )(
\g_5) - G a^- \capc ( \g _\mu \g_5 )( 1) + G a^- \capd ( \g _\mu \g_5
)( \g_5) ],
\label{equ52}
\end{eqnarray}
where, $ a^\pm = \ud (a_L(e)\pm a_R(e)),$ and we have omitted writing
the contractions of the Dirac spinors indices for the initial and
final fermions, respectively.  The photon exchange contribution can be
incorporated by treating the parameters $ a^\pm $ as operators acting
on the vertex functions, $\capa ,\ \capb , \ \capc , \ \capd $, by
means of the formal substitutions,
\begin{eqnarray}
G a^\pm \capa & =& G_Z {a_L(e) \pm a_R(e) \over 2} \bigg ( {A^{JJ'} _L
(f) \pm A^{JJ'} _R (f) \over 2} + a^{JJ'} \bigg ) + G_\g { 2 Q(f)
\choose 0} \bigg ( { A_L^ { \g JJ'} (f) \pm A_R ^{ \g JJ'} (f) \over
2}+ a^{JJ'} \bigg ) , \cr G_Z & =& ({g\over 2 \cos \t_W })^2 {1\over
s-m_Z^2 +im_Z \G_Z} , \ G_\g = ({g\sin \t_W \over 2 })^2 {1\over s}.
\label{equ53}
\end{eqnarray}
Analogous formulas to the above ones hold for the other products, $ G
a^\pm \capb ,\ G a^\pm \capc ,\ G a^\pm \capd.$ We have labelled the
vertex functions for the photon current by the suffix $\g $. The
formulas expressing the RPV one-loop contributions to the vertex
functions are provided in Appendix \ref{appexa}, quoting the results
derived in our previous work. \cite{chemtob} The amplitude $ M^{JJ'}$
in eq.(\ref{equ52}) may be viewed as a $ 4\times 4 $ matrix in the
fermions polarization space, $ (f(p, \l ) \bar f_{J'} (p', \l ') \vert
M \vert l^+ (k',\mu ') l^- (k, \mu ) ) $.  The various products in
eq. (\ref{equ52}) for the matrix elements with respect to the two
pairs of Dirac spinors separate into 8 distinct terms.  The
calculation of the helicity amplitudes is most conveniently performed
with the help of the `Mathematica' package.  Of the $16$
configurations only the $ 8$ helicity off diagonal configurations in
the initial fermions are non vanishing.  The explicit formulas for the
helicity amplitudes are provided in Appendix \ref{appexa}.

\subsection{Charged lepton energy distribution}
\label{sec31}
The differential cross section for top production and decay is
described in the factorization approximation. Ignoring the spin
correlations, which corresponds to dropping the spin non diagonal
contributions between the production and decay stages, yields:
\begin{eqnarray}
d\sigma & =&{\vert p \vert \over 128 \pi s \vert k \vert } {m_t \over
\pi } \int d(\cos \t ) \sum_{ \l } \vert M_{prod} (l^-l^+\to t_\l \bar
c)\vert ^2 \int dp^2 {1\over \vert p^2 -m_t^2 +im_t \G_t\vert ^2 }
d\G_t ,\cr d \G_t& = &{1\over (2\pi )^3 8 m_t }\overline { \sum } _ {
\l ' } \vert M_{dec} (t_{ \l '} \to b l^+ \nu ) \vert ^2 dE ^ \star _l
dE^\star _b .
\label{equ1}
\end{eqnarray}
The production amplitude is denoted, $M_{prod}$, the top decay
amplitude, $ M_{dec}$, and $ \l , \ \l ' = \pm 1 $, are polarization
labels, which will also be written for short as, $\pm $ .  We shall
assume a narrow resonance approximation for the top propagator, $\vert
p^2 -m_t^2 +im_t \G_t\vert ^{-2} \to {\pi \over m_t \G_t } \d (p^2
-m_t^2)$.  For the energies of interest, all the leptons and quarks,
with the exception of the top, may be treated as massless.  Two frames
of interest are the laboratory $(l^-l^+$) rest frame and the top rest
frame.  The letters denoting momentun variables in the $l^-l^+ $
center of mass (laboratory) frame are distinguished from those in the
top rest frame by the addition of a star. Standard kinematical methods
\cite{kajan} can be used to transform variables between these frames.
Exploiting the rotational invariance, one may conveniently choose to
work in the spatial frame where the top momentum lies in the $ xOz $
plane ($\theta , \phi =0 $) and the charged lepton points in an
arbitrary direction described by the spherical angles, $ \t_{l},
\phi_{l}$. The relations between angles may be obtained by use of the
spherical triangle identities, for example, the angle between lepton
and top reads, $ \cos \t_{lt} = \cos \t_{l} \cos \t + \sin \t_{l} \sin
\t \cos \phi_l.$ The Lorentz boost from the top rest frame to the
laboratory frame, involves a velocity parameter, $ \vec v= \vec p/E_p,
\ \beta = p/E_p, \ \g = (1-\b ^2)^{-1/2}= E_p/m_t, $ and yields for
the charged lepton momentum four vector and polar angle relative to
the top momentum, $E^\star _l =\g (E_l - \vec v \cdot \vec k_l ),\quad
\vec k^\star _l= \vec k_l +\g \vec v ({\g \vec v \cdot \vec k_l \over
\g +1} -E_l ), \quad \cos \t_{lt} ^\star ={\cos \t_{lt}-\b \over 1- \b
\cos \t_{lt} }$.

The top differential semileptonic decay rate has been thoroughly
studied in the literature.  \cite{kuhn} One representation convenient
for our purposes is the double differential rate with respect to the
final charged lepton energy, $ E^\star _l$, and the final lepton and
neutrino invariant mass squared, $ W ^2 = (k_l+k_ \nu )^2$.  The
result for the unpolarized rate carries no dependence on the
scattering angles and reads, quoting from ref.\cite{kuhn},
\begin{eqnarray}
d\G _t&= &{N_lG_F^2m_t^5\over 16\pi^3 } dx_l \int dy { x_l
(x_M-x_l)\over (1- y \xi)^2 +\g ^2 }, \cr &=& {N_lG_F^2m_t^5\over
16\pi^3 } {2\over m_t} {x_l (x_M-x_l) \over \g \xi } \tan^{-1} {\g \xi
x_l (x_M-x_l) \over (1+\g ^2) (1-x_l) -\xi x_l (x_M-x_l ) }\
dE_l^\star .
\label{equ4}
\end{eqnarray}
The kinematical variables for the emitted charged lepton and neutrino
are defined as, $ x_l =2E_l^\star /m_t, \ y= W^2/m_t^2, \ [ W=k_l+k_
\nu ] $ with the bounds, $ 0< x_l< x_M , \ 0< y < {x_l (x_M-x_l) \over
1-x_l } $ and we employ the following notations, $ N_l $ for the
number of light lepton flavors, $ \g = \G_W/m_W,\ \xi = m_t^2/m_W^2,\
x_M =1-\e^2, \ \e =m_b/m_t , \ \tan ^{-1} A = Artan \vert A \vert +
\pi \t (-A) $.  Recall that the number of light lepton flavors, $ N_l
$, is set to $ N_l =2$ in our analysis.  A useful trick to obtain the
distribution with respect to the laboratory frame lepton energy, $
E_l$, is to choose the top momentum along the $Oz$ axis fixed frame
and introduce the top rest frame electron energy by means of the
change of variable, $ (E _l , \cos \t^\star _l) \to ( E _l, E ^\star
_l)$, associated with the Lorentz boost between the top rest frame and
the laboratory frame, $ E_l= \g E_l^\star (1+\b \cos \t ^\star _l)
$. The result reads,
\begin{eqnarray}
{d \G_t \over dE_l }=\int_{-1}^{+1} d \cos \t_l^\star {d^2 \G_t \over
dE_l d \cos \t _l ^ \star } = {2\over m_t \g \b } \int_{x_l^-}
^{x_l^+} { d x_l\over x_l } {d ^2 \G _t \over d x_l d \cos \t _l ^
\star } ,
\label{equ41}
\end{eqnarray}
where the integration interval over $ x_l $ is bounded at, $x^\pm _l =
{ 2 E_l \over m_t \g (1\pm \b ) }$.

\subsection{Top polarization observables}
\label{sec32} 
 An essential use will be made of the factorization property of the
double differential distribution for the top decay semileptonic rate
with respect to the emitted lepton energy and angle relative to the
top spin polarization vector. This distribution is described at the
tree level as, $ {d ^2\G_t \over dE_l ^\star d \cos \psi_l } = {d\G_t
\over dE_l ^\star } {1+\cos \psi_l \over 2} $, where, $ \cos \psi_l
=-s(p) \cdot k_l $, is the angle between the lepton momentum and the
top spin polarization vector, $s_\mu (p)$, in the top rest frame.
Equivalently, $ {d ^2 \G _t \over d x_l d \cos \t _l ^ \star } = {d \G
_t \over d x_l } { 1+\cos \psi_l \over 2 } { d\cos \psi_l \over d \cos
\t _l ^ \star }$. \cite{kuhn} As it turns out, this representation
remains valid to a good approximation when one-loop QCD corrections
are included. \cite{kuhn2} We choose to describe the top polarization
in the spin helicity formalism, using techniques familiar from
previous works. \cite{kane,topfor} The definition for the helicity
basis Dirac spinors is provided in Appendix \ref{appexa}.  Since the
polarization axis coincides then with the top momentum, the dependence
on $\psi_l $ can also be simply rewritten as, $ (1+ \cos \psi_l )/2 =(
1 + \l \cos \t_l ^\star )/2 ,$ such that, $ \l = [- 1, +1] $,
correspond to $[L,R]$ helicity, respectively.

The helicity amplitudes associated to the pair of CP-conjugate
processes are related by the action of $ CP$ as, $ <f_{\l }\bar f'_{\l
' } \vert M \vert l^+_{ \mu '} l^- _{ \mu } > \to <f '_{-\l '} \bar
f_{-\l } \vert M \vert l^+_{- \mu } l^- _{- \mu ' } >.$ Unlike the
process, $ l^+ l^- \to t\bar t$, where both the initial and final
states are self-conjugate under CP, here only the initial state is
self-conjugate, while the action of CP relates the different final
states, $ t\bar c $ and $ c \bar t$.  Let us express the amplitudes
for the pair of CP-conjugate processes as sums of tree and loop terms,
$ M^{JJ'} = a_0 +\sum_\a b_\a f_\a(s+i\e ) , \ \bar M^{JJ'} = a^\star
_0 +\sum_\a b^\star _\a f_\a(s+i\e ) $, where the loop terms, $b_\a
f_\a(s+i\e ) $, are linear combinations with real coefficients of the
vertex functions, $ A^{JJ'}_{L}, \ A^{JJ'}_{R}, \ a^{JJ'} , \ i
d^{JJ'} $, with the energy dependent complex functions, $ f_\a (s+i\e
) $, representing the factors in loop amplitudes which include the
absorptive parts.  In terms of these notations, a CP asymmetry
associated with the difference of rates for the pair of CP-conjugate
processes in some given CP-conjugate configurations of the particles
polarizations, can be written schematically as,
\begin{eqnarray} 
\vert <\l \l ' \vert M \vert \mu ' \mu >\vert^2 & - & \vert <-\l ' -\l
\vert M \vert -\mu - \mu ' >\vert^2 \propto \sum_\a Im (a_0 b_{\a
}^\star ) Im (f_\a (s+i\e ) ) \cr & - & \sum _{\a < \a '} Im (b_\a
b_{\a '} ^\star ) Im (f_\a (s+i\e ) f_{\a '} ^\star (s+i\e ) ) .
\end{eqnarray} 
Thus, the necessary conditions for a non vanishing polarized asymmetry
to arise from the tree-loop interference term are a relative complex
CP-odd phase between the tree and loop coupling constants and an
absorptive part from the loop terms.  The angle integrated production
rates for the CP-conjugate reactions, $ l^+ l^- \to t \bar c$ and $
l^+ l^- \to c \bar t$, for the case of polarized top and antitop,
respectively, are obtained by summing over the polarization of the $c,
\ \bar c $ quarks as,
\begin{eqnarray} 
\sigma (t_{L})& =& \sigma (t_L \bar c _R ) + \sigma (t_L \bar c _L ) ,
\ \sigma (t_{R}) = \sigma (t_R \bar c _L) + \sigma (t_R \bar c _R ),
\cr \sigma (\bar t_{L})& =& \sigma (\bar t_L c _R ) + \sigma (\bar t_L
c _L ) , \ \sigma (\bar t_{R}) = \sigma (\bar t_R c _L) + \sigma (\bar
t_R c _R ).
\label{equ54}
\end{eqnarray}
Forming the half differences and sums of rates, $ \d \sigma = \ud (
\sigma (t_L) -\sigma (t_R) ) , \ \d \bar \sigma =\ud ( \sigma (\bar
t_R) -\sigma (\bar t_L) ), \ \sigma _{av} = \ud ( \sigma (t_L) +\sigma
(t_R) ), \ \bar \sigma _{av}= \ud ( \sigma (\bar t_R) +\sigma (\bar
t_L) )],$ such that, $\sigma (t_{L, R } ) = \sigma _{av} \pm \d \sigma
, \ \bar \sigma (t_{ R, L }) =\bar \sigma _{av} \pm \d \bar \sigma ,$
one can define the following two CP-odd combinations,
\begin{eqnarray}
\cala = {\sigma _{av} -\bar \sigma _{av} \over \sigma _{av} +\bar
\sigma _{av} }= {\sigma _{t\bar c} - \sigma _{\bar t c} \over \sigma
_{t\bar c} + \sigma _{\bar t c} } , \quad \cala ^{pol} & =& { \d
\sigma -\d \bar \sigma \over \sigma _{av} + \bar \sigma _{av} },
\label{equ55}
\end{eqnarray}
which will be designated as unpolarized and polarized integrated rate
asymmetries.  The above definition for the unpolarized asymmetry, $
\cala $, is identical to the one studied in our previous
work. \cite{chemtob} The asymmetries depend on the RPV coupling
constants through the ratio of loop to tree amplitudes as, $ Im ( { \l
^{'\star } _{iJk } \l '_{iJ'k} \over \l ^{'\star } _{1Jk'} \l
'_{1J'k'} } ) \propto \sin \psi $, where the dependence on the CP
violation angle parameter, $\psi $, reflects the particular
prescription adopted in this study to include the CP-odd phase.  The
index $ k ' $ refers to the d-squark generation in the tree amplitude
and the indices $ i, k $ to the fermion-sfermion generations for the
internal fermion-sfermion pairs, $ {d_k \choose \tilde e^\star _{iL}}
, \ {e^c_i \choose \tilde d_{kR}} $, in the loop amplitude.

  It is important not to confuse the above analysis with that of the
top-antitop pair production, $ l^- l^+ \to t\bar t$, where a CP-odd
asymmetry observable for a single final state may be defined in terms
of the difference of helicity configurations, $ \s (t_L \bar t_L ) -
\s (t_R \bar t_R ) $.  A non vanishing value for the corresponding
difference of polarized rates can only arise via tree-loop
interference terms involving the absorptive part of the top quark
electric dipole moment, $ Im (d^{JJ}) $. \cite{schmidt,chang1} One
should note here that the one-loop contribution of the RPV $\l '$
interactions to $ Im (d_t ^{JJ}) $ vanishes.  Two closely related
processes, which are amenable to an analogous treatment, are the
$b\bar b$ quark pair \cite{valencia} and $\tau ^+ \ \tau ^-$ lepton
pair production.  Double spin correlation observables for the latter
reaction, $ l^- l^+ \to \tau ^- \tau ^+ $, have been examined in a
recent work.  \cite{shalom} We note that the RPV $\l $ interactions
can give a non vanishing contribution to $ Im (d_\tau ^{JJ}) $.

\begin{figure} [h]
\centerline{ \psfig{figure=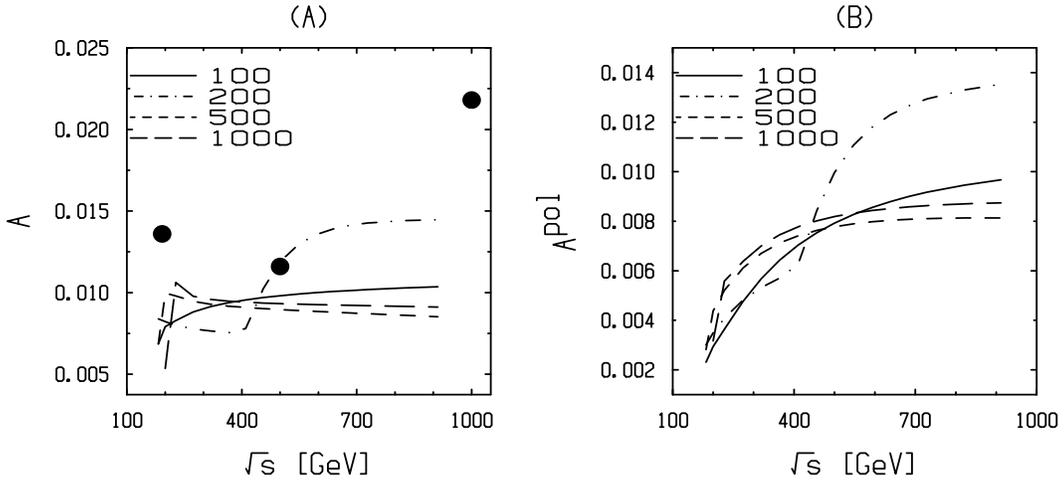,width=14cm} }
\vskip 1 cm
\caption{ The CP-odd production rate asymmetries as a function of the
center of mass energy, $ s^\ud $, for fixed values of the down squark
mass, $\tilde m = [100, \ 200, \ 500 , \ 1000 ] \ GeV$.  The left hand
plot $ (A)$ gives the unpolarized asymmetry, $ \cala = ( \sigma _{av}
- \bar \sigma _{av} ) /(\sigma _{av} + \bar \sigma _{av} )$.  The
upper bounds for the absolute values of the statistical errors on the
asymmetries, as evaluated with $\l '_{12k}\l '_{13k} =0.1, \ \tilde m
= 100 \ GeV$ and integrated luminosities $\call =100.  \ fb^{-1}$ are
shown as full circles.  The right hand plot $ (B) $ gives the spin
polarization dependent asymmetry, $\cala ^{pol} = (\d \sigma -\d \bar
\sigma ) /(\sigma _{av} + \bar \sigma _{av} ) $. }
\label{figbspol}
\end{figure}

The results for the rate asymmetries are displayed in
Fig.\ref{figbspol}.  The numerical results for the unpolarized case
(window $(A)$ in Fig.\ref{figbspol}) update the results presented in
ref. \cite{chemtob} since the present calculation includes the
contributions from the Lorentz covariant tensorial ($\s _{\mu \nu }$)
coupling which were ignored in our previous work.  \cite{chemtob} The
asymmetry for the polarized case (window $(B)$ in Fig.\ref{figbspol})
involves the difference of the spin helicity asymmetry in the total
production cross sections for the CP mirror conjugate top and antitop
mirror reactions. While this CP-odd polarized asymmetry is not
directly observable, it enters as an important intermediate quantity
in evaluating the measurable kinematic distributions of the top decay
products dependent on the top spin.  We have assumed all the relevant
RPV coupling constants to be equal and set the CP-odd phase at $\sin
\psi =1$. The rapid change in slope for the $\tilde m =200\ GeV$ case
are due to the threshold effect from the imaginary part in the
superpartner one-loop contributions, which set at $ \sqrt s = 400 \
GeV$.  Aside from this large discontinuous contribution, one sees that
both asymmetries comprise another contribution which is nearly
independent of $\tilde m$ and increases smoothly with the initial
energy.  Both asymmetries, $ \cala $ and $ \cala^{pol} $, take values
of order a few $ 10^{-3} $, reaching $ O(10^{-2}) $ at the highest
incident energies.

The statistical uncertainties on the asymmetry may be evaluated in
terms of the signal cross sections and the integrated luminosity by
considering the approximate definition, $ \d \cala = 1/ [\call (\s
_{t\bar c} + \s _{\bar t c})] ^\ud $.  Using the same input value for
the luminosity $ \call = 100.  \ fb ^{-1} $ at the three cm energies,
$\sqrt s =[0.192, \ 0.5 , \ 1. ] \ TeV$, along with the cut signal
rates in Table \ref{table1}, we obtain statistical errors on the
asymmetries of order $ O(10^{-1})$. These values lie nearly two order
of magnitudes above the value obtained for the signal. At this point,
it is important to observe that in getting the above estimates for the
rates we have been using somewhat conservative assignments for the RPV
coupling constants.  As already noted, the single top production cross
sections could possibly be two order of magnitudes larger if we were
to use coupling constants values of order, $\l '_{12k} \l '_{13k}
\simeq 10^{-1}$.  Such values are compatible with the indirect bounds
only for the extreme down squark mass $ \tilde m = O(1\ TeV)$ range.
In the hypothetical case where the production rates would be enhanced
by two order of magnitudes, the statistical errors on the asymmetries
would correspondingly get reduced by a factor $ O(10^{-1})$, thereby
reaching the same order of magnitude as the signal asymmetries.
Nevertheless, as plotted in window $(A)$ of Fig.\ref{figbspol}, the
corresponding errors would still be somewhat larger than the signals.
We should note here that the contribution to the one-loop amplitude
from internal sfermion and fermion lines belonging to the third
generation is controlled by the coupling constants quadratic product,
$ \l '_{323} \l '_{333} $, which is subject to weak constraints.
Should the RPV coupling constants exhibit a hierarchical structure
with respect to the quarks and leptons generations, one cannot exclude
the possibility of a factor $10$ enhancement from the ratio, $ Im ( \l
^{'\star }_{323} \l '_{333} / \l ^{'\star } _{123} \l '_{133} )$.
Such an order of magnitude gain on this ratio would raise the
asymmetries up to $ O(10^{-1})$ bringing them well above the
experimental uncertainties.  Lastly, we observe that a more complete
formula for the uncertainties on the asymmetries reads, $ (\d \cala
)^2 = 2 (\d \s_{t\bar c})^2 [1- C +(1+C) \cala ^2 ]/ (\s _{t\bar c} +
\s _{\bar t c})^2 $, where we used equal standard deviations for the
CP conjugate reactions rates, $ \d \s_{t\bar c} = \d \s_{\bar t c} $,
and denoted the correlated error on these two rates as, $ C = <\d
\s_{t\bar c} \d \s_{\bar t c}> / \d \s_{t\bar c}^2 $.  Clearly, an
improvement on the statistical treatment of the $ t\bar c +\bar t c$
events sample, allowing for a positive non vanishing value of the
error correlation associated with the identification of isolated
single negatively and positively charged lepton events, should greatly
help in reducing the experimental uncertainties caused by the small
event rates.

\begin{figure} [t]
\centerline{ \psfig{figure=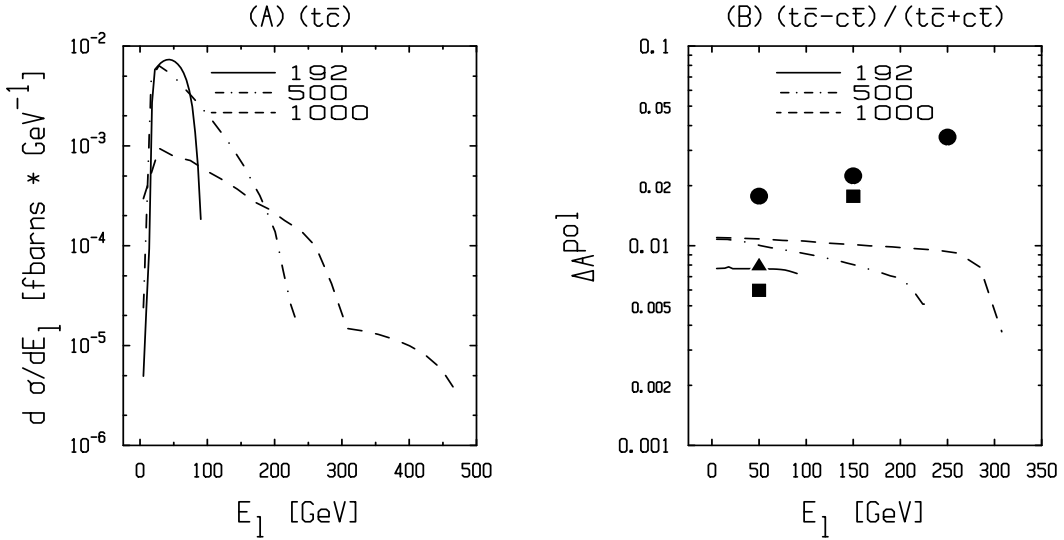,width=14cm} }
\vskip 1 cm
\caption{Energy distribution for the charged lepton as a function of
the laboratory frame lepton energy, for a set of center of mass
energy, $ s^\ud =[ 192, \ 500 , \ 1000] \ GeV$.  The parameters are
set at, $ \l ' =0.1, \ \tilde m =100 \ GeV$.  The left hand plot $ (A)
$ gives the differential lepton energy distribution, $ { d \sigma
\over dE_l }$. The right hand plot $ (B) $ gives the asymmetry in the
energy distribution for leptons of opposite charges in the
CP-conjugate final state channels, $(t\bar c) $ and $ \ (c\bar t)$: $
\D \cala ^{pol}= [ {d \sigma ^+ \over d E_l} - {d \sigma ^- \over d
E_l}]/ [{d \sigma ^+ \over d E_l} + {d \sigma ^- \over d E_l} ].$ The
upper bounds for the absolute values of the statistical errors on the
asymmetries, as evaluated with $\l '_{12k}\l '_{13k} =0.1$ and with
integrated luminosities, $\call =100. \ fb^{-1}$, are shown for three
energy bins of width $ 100 \ GeV$ each, centered at the charged lepton
energies, $ E_l= (50,\ 150,\ 250 ) \ GeV $.  The results for three
values of the center of mass energy, $ s^\ud =[ 192, \ 500 , \ 1000] \
GeV$ are displayed by full triangles, squares and circles. }

\label{figenasy}
\end{figure}

The energy distribution for the negatively and positively charged
leptons in the pair of CP-conjugate reactions may be defined as,
\begin{eqnarray}
<\s ^+> \equiv <{ d \sigma ^+ \over dE_l }> = < \sigma (t_L) f_L +
 \sigma (t_R) f_R > , \quad <\sigma ^- > \equiv < { d \sigma ^- \over
 dE_l }> = < \sigma ( \bar t_R) f_L + \sigma (\bar t_L) f_R > ,
\label{equ56}
\end{eqnarray}
where the correlations between the top spin lepton momentum are
described by the factors, $ f_{L,R}= \ud (1\mp \cos \t^\star _l )$,
and the brackets stand for the angular integration.  The occurrence of
angular correlation factors of opposite signs in the $\bar t$
production case accounts for the kinematical fact that the antitop is
oriented in space with a momentum $ - \vec p$. A CP-odd charge
asymmetry observable with respect to the charged lepton energy
distribution may be defined by considering the following normalized
difference of distributions,
\begin{eqnarray}
\D \cala ^{pol} & =& { <\s ^+ > - <\s ^- > \over <\s ^+ > + <\s ^- > }
={ (\sigma _{av} -\bar \sigma _{av}) +< (\d \sigma - \d \bar \sigma )
(f_L -f_R) > \over (\sigma _{av} +\bar \sigma _{av}) + < (\d \sigma +
\d \bar \sigma ) (f_L -f_R) > }.
\label{equ57}
\end{eqnarray}
The numerical results for the charged lepton energy distributions and
for the above defined charge asymmetry in the lepton energy
distributions are displayed in Fig.\ref{figenasy}.  (Note that the
transverse energy distribution, in the plane orthogonal with respect
to the top momentum, may be simply obtained as, ${ d\G \over dE_{lT}}
= { d\G \over dE_{l}} {1\over \sin \t_{lt} ^\star }$.  The
distribution in the plane orthogonal to the collision axis is less
trivial to evaluate since this requires an additional integration over
the lepton azimuthal angle.)  The energy distributions for the
unpolarized cross section essentially reproduce the results found in
our above quoted event generator predictions, Fig.\ref{figtopp}.  The
energy distributions for the polarized asymmetry lie at values of
order of magnitude, $O( 10^{-3}) $, always retaining the same positive
sign as the lepton energy varies.  For a fixed energy of the emitted
lepton, the asymmetry increases with the initial energy, reaching
values of order $O( 10^{-2}) $.  In window $(B)$ of Fig.\ref{figenasy}
we have plotted the experimental uncertainties using the same inputs
for the luminosities and the rates as in the discussion of the
unpolarized asymmetries given above.  To ease the comparison with
experiment, we divide the charged leptons energy interval into three
bins of width $ 100 GeV$ each, centered at the three lepton energies,
$ E_l= (50,\ 150,\ 250 ) \ GeV $.  The statistical errors on the
asymmetries in the energy distributions lie at the same level as those
associated to the total asymmetries, so that similar conclusions
should apply.  Setting ourselves within the same optimistic scenario
by using $\l '_{12k} \l '_{13k} =10^{-1}$ and $\call= 100 fb^{-1}$, we
obtain expected errors of order $ O(10^{-2})$. These values are
insufficient for a comfortable identification of a signal asymmetry.
However, we reiterate, as in the above discussion, that an enhancement
of the signal asymmetries to an observable level of $ O(10^{-1})$, due
to a hierarchical structure in the generation dependence of the $\l
'_{ijk}$, is a real possibility.

\section{Conclusions}
\label{sec4}
We have demonstrated that single top production through the RPV
interactions could be observed at the future linear colliders or else
be used to set bounds on the RPV coupling constants, $ \l ' _{12k} \l
'_{13k} < O(10^{-2})$, over a wide interval for the down squark mass,
$ m _{\tilde d_{kR} } < 1. \ TeV$.  The $ b $ quark tagging would help
greatly to overcome the background. Even with an imperfect $ b $ quark
tagging, it is still possible to drastically reduce the background,
from $ WW $ and $ b\bar b$, without much harming the signal. The
analysis of top polarization observables via the semileptonic decay
channel of the top allows to test for the presence of a CP violating
complex phase, embedded in quadratic products of the RPV coupling
constants. We have focused on the asymmetry in the energy
distributions of the charged leptons in the CP-conjugate pair of final
states, $b l^+ \nu \bar c$ and $\bar b l^- \bar \nu c $, obtaining
asymmetries of order $ 10^{-3} \ - \ 10^{-2} $ for the incident
energies expected at the future leptonic colliders.  These values lie
somewhat below the anticipated limits of observability.  However, it
may be possible to obtain enhanced values of order $ 10^{-1}$, should
the RPV coupling constants $\l'_{ijk}$ exhibit large hierarchies with
respect to the quarks or leptons generations.  Future promising
extensions might include analogous reactions accessible with
lepton-photon or photon-photon colliding beams, $ l \g \to t\bar c , \
\g \g \to t\bar c $, where the expected production rates are
substantially larger than those for the $ l^-l^+ $ colliders.

\appendix
\section{}
\setcounter{subsection}{0} \setcounter{equation}{0}
\renewcommand{\theequation}{A.\arabic{equation}}
\renewcommand{\thesubsection}{A.\arabic{subsection}}
\label{appexa}
{\bf Helicity amplitudes.}

 The helicity spin basis Dirac spinors for a fermion or an
antifermion, of mass $ m$ and four momentum, $ k_\mu = (E_k = (k^2
+m^2) ^\ud, \vec k)$, and polar coordinates, $ \vec k = (\t , \phi )$,
can be written in the form of direct products of the Dirac spinor
two-component space with the the two-component space of Pauli helicity
basis spinors, $ \phi _\l (\vec k) , $ satisfying, $ \vec \s \cdot
\hat k \phi _\l (\vec k) = \l \phi _\l (\vec k). $ In the Dirac
representation for the Dirac matrices, $\g_0 = \b ,\ \vec \g = \b \vec
\a ,\ \g_5 = \pmatrix{0 & 1 \cr -1 & 0}$, the spinors read,
\begin{eqnarray}
u(\vec k, \l )&=& \sqrt {\e_k }\pmatrix {1\cr \tilde k \l } \times
\phi _\l (\vec k ), \ v(\vec k, \l )=\sqrt { \e_k } \pmatrix {- \tilde
k \l \cr 1 } \times \phi_{-\l } (\vec k) ,\ \cr \phi _{-1}(\vec k) &
=& \pmatrix{-\sin (\t /2) e^{-i\phi } \cr \cos (\t /2) } , \phi
_{+1}(\vec k) = \pmatrix { \cos (\t /2) \cr \sin (\t /2) e^{+i\phi }
},
\label{equ5}
\end{eqnarray}
where, $\e_k= E_k+m, \ \tilde k= \vert \vec k \vert /(E_k+m), $ and $
\chi_{\l }, \ [\l =\pm 1] $, are the Pauli spinors in the basis with a
fixed quantization axis identified with the spatial three-axis, $Oz$.
The helicity basis spin eigenstates with a space parity reversed three
momentum are defined as, $\phi_{\l } (-\vec k) = e^{ -i (\phi + \pi )
(\l ' -\l )/2 } (e^{-i{(\pi -\t ) \over 2} \s_y })_{\l '\l } \chi_{\l
' } = \phi_{\l } (\vec k)\vert _{[\t \to \pi -\t , \ \phi \to \phi
+\pi ]} $.

The $8$ non vanishing helicity amplitudes for the process, $ l^+(k' ,
\mu ') + l^- (k, \mu) \to u_J (p, \l ) +\bar u_{J'} (p', \l ') $, are
listed in the formulas below:
\begin{eqnarray} M_1= M(+-+-) & = & 4 \calf \, [ - ( ( 1 + {\tilde
p}\, {\tilde p'} ) \, ( {\ya} + {\yb} ) \ + ( {\tilde p} + {\tilde p'}
) \, ( {\yc} + {\yd} ) ] \, {{\sin ^2({{\t } / {2}})}}, \cr M_2= M
(+-++) & = & 2 \calf \, [ ( -1 + {\tilde p}\, {\tilde p'} ) \, (
{\ya}+ {\yb}) + ( {\tilde p} - {\tilde p'} \ ) \, {\yc} + ( {\tilde p}
- {\tilde p'} \ ) \, {\yd} \cr & + & 2\,p\, ( {\tilde p} + {\tilde p'}
) \, ( {\ye} + {\yf}) + 2\,p\, ( 1 + {\tilde p}\, {\tilde p'} ) \,(
{\yg} + {\yh}) ] \,\sin (\t ) ,\cr M_3= M (-++-) & = & -4 \calf \, [ -
( 1 + {\tilde p}\, {\tilde p'} \ ) \, ( -\ya + {\yb} ) - ( {\tilde p}
+ {\tilde p'} \ ) \, ( {\yc} - {\yd} ) ] \, {{\cos^2 ({ {\t } / {2}})}
} ,\cr M_4 = M (-+++) & = & 2 \calf \, [ ( -1 + {\tilde p}\, {\tilde
p'} \ ) \,(-\ya + {\yb} ) + ( - {\tilde p} + {\tilde p'} \ ) \, (
{\yc} - {\yd}) \cr & -& 2\,p\, ( {\tilde p} + {\tilde p'} ) \, ( {\ye}
- {\yf}) - 2\,p\, ( 1 + {\tilde p}\, {\tilde p'} ) \, ( {\yg} - {\yh}
] \,\sin (\t ) ,\cr M_5 = M (+---)& = & 2 \calf \, [ ( -1 + {\tilde
p}\, {\tilde p'} ) \, ( {\ya} + {\yb}) + ( - {\tilde p} + {\tilde p'}
\ ) \, {\yc} + ( - {\tilde p} + {\tilde p'} \ ) \, {\yd} \cr & + &
2\,p\, ( {\tilde p} + {\tilde p'} ) \, ( {\ye} + {\yf}) - 2\,p\, ( 1 +
{\tilde p}\, {\tilde p'} ) \, ( {\yg} - {\yh}) ] \,\sin (\t ) ,\cr
M_6= M (+--+) & = & -4 \calf \, [ (1+ {\tilde p}\, {\tilde p'} ) \, (
{\ya} + {\yb}) + ( {\tilde p} + {\tilde p'} ) \, ( {\yc} + {\yd} ) ]
\, {{\cos ^2({{\t } / {2}})} } ,\cr M_7= M (-+--) & = & 2 \calf \, [ (
-1 + {\tilde p}\, {\tilde p'} \ ) \,( - {\ya} + {\yb} ) + ( {\tilde p}
- {\tilde p'} \ ) \, ( {\yc} - {\yd}) \cr & - & 2\,p\, ( {\tilde p} +
{\tilde p'} ) \,( {\ye} - {\yf}) + 2\,p\, ( 1 + {\tilde p}\, {\tilde
p'} ) \, ( {\yg} - {\yh} ) ] \,\sin (\t ) ,\cr M_8= M ( -+-+) & = & 4
\calf \, [ - ( 1 + {\tilde p}\, {\tilde p'} ) \, ( {\ya} - {\yb} ) - (
{\tilde p} + {\tilde p'} ) \, ( {\yc} - {\yd} ) ] \, {{\sin ^2({{\t }
/ {2}})}}.
\label{equ7}
\end{eqnarray}
The arguments refer to the fermions helicity in the following order, $
M_i((h_{e^+},h_{e^-} ,h_f ,h_{\bar f} ) $.  The remaining helicity
amplitudes, omitted from the above list, are understood to vanish
identically. We denote by $\t $ the top scattering angle, $\cos \t =
\vec k \cdot \vec p $, by $ [E_p, E '_{p }] =(s\pm m_J^2 \mp m_{J'}^2
)/2\sqrt s $, the top and charm quarks energies, and use the following
abbreviated notations, $ \tilde p = {p\over E_p +m_J }, \ \tilde p ' =
{p\over E '_{p } +m_{J'} }, \ \calf = {1\over 2} [s (E_p +m_J) (E
'_{p} + m_{J'} )] ^\ud , $ along with the useful compact notations,
\begin{eqnarray}
X_1& = & G a^ + \capa +\calr, \ X_2= G a^- \capa +\calr, \ X_3 = G a^+
\capb +\calr ,\ X_4 = G a^- \capb + \calr , \ \cr X_5& =& \ud G a^+
\capc , \ X_6= \ud G a^- \capc , \ X_7 = \ud G a^+ \capd , \ X_8=\ud G
a^- \capd ,
\label{equ7p}
\end{eqnarray}
where $ G a^\pm \capa , \cdots $ are defined in eq. (\ref{equ53}),
$\calr $ in eq. (\ref{equ323}), and $\capa , \cdots , \capd $, in
eq. (\ref{equ51}).

{\bf One-loop RPV vector boson vertex functions.}

 The one-loop vertex functions, as derived in \cite{chemtob}, are
given by the formulas,
\begin{eqnarray}
&& A_L^{JJ' } = \FLU [ {a_L(u) }\, {B_1^{(2)}} + {a(f_L)}\, {{
{m_f}}^2} C_0 + {a(\tilde f')} ( 2 {\tilde C_{24}} + 2 \ m_J^2\, (
{\tilde C_{12}}\, - {\tilde C_{21}}\, + {\tilde C_{23}}\, - {\tilde
C_{11}}\, ) ) \cr &+& {a(f_R)}\, ( {B_0^{(1)}} - 2\, {C_{24}} - {{
{m_{\tilde f'}}}^2} {C_0} + {{ {m_{J}}}^2} ( C_0 +3 C_{11} -2C_{12}
+2C_{21} -2C_{23} ) -m_{J'}^2 C_{12} ) ], \cr && A_R^{JJ'} = \FLU m_J
m_{J'} [ 2 {a(\tilde f')}\, ( - {\tilde C_{23}} + {\tilde C_{22}} ) +
{a(f_R)}\, ( -C_{11} +C_{12} -2C_{23} +2 {C_{22}} ) \, ] \cr && {
a^{JJ'}\choose - id^{JJ'} } = \FLU {m_J +m_{J'} \over 4}\bigg [ \pm
m_J [ a(f_R) (C_{11}-C_{12}+C_{21}-C_{23}) \cr & -& a(\tilde f ')
(\tilde C_{11}+\tilde C_{21}-\tilde C_{12}-\tilde C_{23}) ] + m_{J'}
[a(f_R) (C_{22}-C_{23}) +a(\tilde f ') (\tilde C_{23}-\tilde C_{22}) ]
\bigg ].
\label{eqm11}
\end{eqnarray} 
The relevant configurations for the internal fermion and sfermion
propagating in the loop are: $ {f \choose \tilde f'} = {d_k \choose
\tilde e^\star _{iL}} , \ {e^c_i \choose \tilde d_{kR}} $.  The
notations for the Passarino-Veltman two-point and three-point
integrals, as specified in our work, \cite{chemtob} are defined
according to the following conventions, $B_A^{(1)}= B_A(-p-p', m_f,
m_f), \ B_A^{(2)}= B_A(-p , m_f, m_{\tilde f'}), \ [A =0,1] $ and $C_A
= C_A (-p,-p',m_f,m_{\tilde f'} , m_f) , \tilde C_A = C_A
(-p,-p',m_{\tilde f'}, m_f,m_{\tilde f'}) .  \ [ A= 0, 11,
12,21,22,23]$ The integral functions with a tilde are associated with
the one-loop diagram for the sfermion current.

\end{document}